  \documentstyle[eqsecnum,aps,preprint,epsf]{revtex}

  \newcommand{\be}{\begin{eqnarray}}
\newcommand{\ee}{\end{eqnarray}}
\begin{document}
 \thispagestyle{empty}
  \title{On the effects of the final state interaction in the
  electro-disintegration
  of the deuteron at intermediate and high energies}
  \author{ C. Ciofi degli Atti,  L.P.  Kaptari
\thanks{On leave from Bogoliubov Lab. Theor. Phys., JINR, Dubna,
Russia.}}
 \address{  Department of Physics, University of Perugia, and
 INFN, Sezione di Perugia,
 via A. Pascoli, Perugia, I-06100, Italy}
  \author{ D. Treleani}\address{Department of Theoretical Physics,
  University of Trieste, Strada Costiera 11, INFN, Sezione di
  Trieste and ICTP, I-34014 Trieste, Italy}

 \date{\today}
 \maketitle
 \begin{abstract}
The role of the  final state interactions (FSI) in the
inclusive  quasi-elastic
disintegration of the deuteron is investigated
treating  the two-nucleon  final state within the exact continuum
solutions of the non-relativistic Schroedinger equation, as well as within  the
Glauber multiple scattering approach. It is shown that for values
of the Bjorken scaling variable $x_{Bj}\simeq 1$ both approaches
provide similar results, unless the case   $x_{Bj}\gtrsim 1$,
where they appreciably disagree. It is demonstrated that present
experimental data, which are mostly   limited to a region of four-momentum transfer
($Q^2 \lesssim 4 (GeV/c)^2$)  where  the
Center-of-Mass energy of the final state is below the pion threshold
production, can be satisfactorily reproduced by the approach based on the
exact solution of the Schroedinger equation and not by the Glauber
approach. It is also pointed out that the latter, unlike the
former, does not satisfy the  inelastic Coulomb sum rule, the violation
being of the order of
 about 20\%.
 \end{abstract}

\section{INTRODUCTION}
\label{sec:introduction}
The role played by  the effects of final state interaction (FSI) in
electro-disintegration processes
 is
a very  relevant issue, for they may in principle hinder
the extraction of reliable information  not only on
nuclear structure, but also on fundamental hadronic properties
in the  medium, which could be obtained
from different kinds of lepton scattering processes off nuclear
targets.
  Apart from the
few-body systems at low energies, for which exact
solutions of the Schroedinger equation in the continuum are
becoming to be  available (see e.g.
~\cite{gloeckle,rosati}), the treatment of  FSI effects in
complex
systems at   intermediate and high
energies still requires the use of several approximations.
This concerns both the  semi-inclusive processes
  $A(e,e'p)X$
   (see e.g. \cite{semiincl}),
   and  the fully inclusive
process $A(e,e')X$,
 for which
several methods have been proposed with conflicting results (see e.g.
~\cite{incl}).
 Most of these
 approaches
  rely on the use of
the Glauber multiple scattering theory, assuming that
the struck nucleon, after $\gamma^*$ absorption, is
on shell and  propagates in the medium
with total energy  $\sqrt {({\bf q}+ {\bf p})^2 +M^2}
\simeq \sqrt {{\bf q}^2 +M^2}$ ($\bf q$
and $\bf p$ are the three-momentum transfer and
the momentum of the struck nucleon before
interaction, respectively). The latter assumption, which is
a  very reasonable one
at   $x_{Bj} \simeq 1$
($x_{Bj}=Q^2/(2M\nu)$  is  the Bjorken scaling
variable ~\cite{Bjork},   $Q^2={\bf q}^2 -{\nu}^2$   the four-momentum
transfer,
and $M$ the nucleon mass), could
be questionable at higher or lower values of
$x_{Bj}$,  where the struck nucleon, after $\gamma^*$ absorption,
is far off-shell; moreover,  even at high values of
$|{\bf q}|$,  the two nucleon relative energy  might be  not
sufficiently high to justify the use of the
Glauber high energy approximation, so that
 a careful
consideration of the two-nucleon kinematics is called for.
As a matter of fact, it has been
shown ~\cite{ciofideut}
that  existing data on the inclusive electro-disintegration
of the deuteron, $D(e,e')X$ \cite{Bosted},
correspond, at  $x_{Bj}> 1$,  to a very low relative energy of the
two nucleon final state even
if $|\bf q|$ is very large, and that they can be
 satisfactorily  explained by using for the continuum
 state the solution of the non relativistic Schroedinger equation \footnote{From
 now on,  the method based upon the exact solution of
 the non-relativistic Schroedinger equation to generate bound and
 continuum two-nucleon states, will be referred to a as the \it{Schroedinger approach}}
  . It is however clear that, given   a
   fixed value of  $x_{Bj}$, if
$|\bf q|$ (i.e. $Q^2$) is further increased, inelastic processes
 could become operative
 and the
Schroedinger approach  becomes  inadequate.
 Within these  kinematical
conditions,   i.e.  at high
relative energies of the $np$-pair in the continuum,
 the Glauber
approach  has been frequently used to calculate FSI effects,
 which, however, requires
several approximations in case of complex nuclei. In the
deuteron case,  FSI  effects can be calculated exactly within
both the Schroedinger and the Glauber approaches.
It is just the aim of this paper to present  the results of
such a  calculation  for    the inclusive  electro-disintegration
of the deuteron $D(e,e')X$ in the quasi-elastic
region,  i.e. at
$\nu \le Q^2/2M $, or  $x_{Bj}>1$. In order  to better
display the effects of the FSI,  our results will be presented
not only in terms of cross sections, but also in terms of y-scaling
functions\cite{ciofideut}. Our paper is organized as follows: in Chapter II the
basic formalism of inclusive processes within the Plane Wave Impulse
Approximation (PWIA)  is recalled; the formalism pertaining to the
treatment of the
FSI within the Schroedinger and the Glauber approaches is illustrated
in Chapter III; the results of calculations are given in Chapter IV;
the Conclusions are drawn in Chapter V.

\section{The One Photon Exchange and the Plane Wave Impulse Approximations}
\label{sec:Basic}
In this Section the relevant formulae describing  the inclusive cross section
 $D(e,e')X$
 will be recalled. In the One Photon Exchange Approximation,
 depicted in fig.\ref{pict1},
 the inclusive cross section reads as follows
 \begin{equation}
 \frac{d^3\sigma}{d\Omega'd{\cal E}'}
 =\sum\limits_f |< \textbf{P}_f,f|\, \hat O\, |i,\textbf{P}_i>|^{\,2}\,
 \delta (\nu +\varepsilon_i -\varepsilon_f),
 \label{eq1}
 \end{equation}
 where $|i>$ and  $|f>$ are the initial and final eigenfunctions of the
 intrinsic nuclear Hamiltonian,
 $\hat O = K\cdot j_\mu\displaystyle\frac{1}{Q^2}J^\mu$,   $j_\mu$ and
 $J_\mu$ are  the electromagnetic currents of the electron
 and the deuteron,
 respectively, and $K$ is  a kinematical factor (see below).

 The 4-momenta of the initial and final
electrons in the laboratory system are $k=({\cal
E},{\bf  k})$ and $k'=({\cal E'},{\bf  k'})$,
respectively,
 the four momentum transfer is  $q=k-k'=(\nu, {\bf  q})$, and
the orientation of the coordinate system is
defined by  ${\bf q}=(0,0,q_z)$.

 At high energies
 the electron mass can be disregarded, so that
\begin{eqnarray}
k^2 = (k')^2 \simeq 0,\quad kk'= -kq =
\frac{-q^2}{2} = \frac{Q^2}{2}, \label{kin10}\\
Q^2 \equiv - q^2 = 4{\cal E
E}'\sin^2{\frac{\theta}{2}}.\label{kin11}
\end{eqnarray}
where $\theta$ is the scattering angle. The
following relations will  be used in what
follows:
\begin{eqnarray}
&& {\cal E} = \frac{\nu}{2}\left( 1+ \frac{
                       \sqrt{\sin^2{\frac{\theta}{2}}+\frac{Q^2}{\nu^2}}
                                        }{\sin{\frac{\theta}{2}}}\right),\quad
                                      {\cal E}' = {\cal E} -\nu \label{kin20}\\[1mm]
&& \left | {\bf  q}\right| = \left | q_z\right|
=\sqrt{Q^2+\nu^2}. \label{kin22}
\end{eqnarray}

In  PWIA, depicted in fig. \ref{pict2},
the three-nucleon momenta in the deuteron,
before interaction, are  ${\bf p}_1=-{\bf p}_2$ and, after
interaction, ${\bf p}{_1}'={\bf q}+ {\bf p}_1$ and  ${\bf p}{_2}'= {\bf
p}_2$;
the relative and center of mass (CM) momenta are
${\bf p}=(1/2)({\bf p}_1- {\bf p}_2)={\bf p}_1$ and
${\bf P}=({\bf p}_1+ {\bf p}_2)=0$.
 The PWIA cross section  in the lab system has the following form
($\bf  p_1 = -{\bf p_2}$):
\begin{eqnarray}
&& {d^3 \sigma \over  d \Omega ' d {\cal E}'
}=\int \sigma_{Mott} \ \sum\limits_{N_i=1,2}\left
[ V_L|\langle p_1|\hat
J_L^{N_i}(Q^2)|p_1'\rangle|^2+ V_T|\langle
p_1|\hat J_T^{N_i}(Q^2)|P_1'\rangle|^2 \right
]\times\nonumber
\\&&\phantom{{d^3 \sigma \over  d \Omega  d {\cal E}' }=\int}
\left [ \frac{M^2   d {\bf p}_2 }{E_1'E_2} \delta
(M_D+\nu -E_1'-E_2) \right ]\ n(|{\bf p}|),
\label{pwia}
\end{eqnarray}
where $L(T)$ refer to the longitudinal (transverse) part of the
nucleon current operator,
$V_{L(T)}$ are the corresponding well-known kinematical factors
$(V_L = \displaystyle\frac{Q^4}{ |{\bf q}|^4},
 V_T = \tan^2(\theta / 2) + \displaystyle\frac{Q^2}{ 2|{\bf q}|^2})$,
 and
 $ n^D(|{\bf p|})$ is the nucleon momentum distribution in the deuteron
\begin{equation}
 n^{D}(|{\bf p}|) =\frac{1}{3(2\pi)^3} \sum\limits_{{\cal M}_D}
\left |\int \Psi_{{1,\cal M}_D} ({\bf r})\ \exp (i{\bf
p}{\bf r}) d{\bf r} \right |^2, \label{deutwf}
\end{equation}
where  $ \Psi_{1,\cal M_D}^D (\textbf{r})$ is  the non relativistic
deuteron wave function, with the two nucleon relative co-ordinate given
by $\textbf{r}=\textbf{r}_1 - \textbf{r}_2$.
 It
is a common practice to express the cross section
(\ref{pwia})  in terms of the free electron-nucleon cross
section for  an on mass-shell nucleon, i.e.
to extrapolate the Rosenbluth cross section to
the off-mass shell case \cite{forest}.
 Since  energy conservation in the two cases  is different (whereas
 the three momentum conservation
 is the same) the  extrapolation unavoidably  requires additional,
 {\it ad hoc} assumptions. In this paper we adopt the
  prescription
of  \cite{forest}, according to which
 the hit nucleon is considered to be on-shell,
 i.e. with a four momentum equal to the one of a free nucleon, {\it
 {\it viz.}}
 $p_1 ^ {on} = ( \sqrt{ {\bf p}_1^2   + {M}^2 }, {\bf p}_1)$,
  and in (\ref{pwia}) the replacement
$\nu  \longrightarrow \bar\nu= \nu + M_D - \sqrt
{ M ^2 + {\bf p}_1^2 } - \sqrt{  M^2+{\bf p}_2^2
}$
 is done, so that $\delta (M_D + \nu - E_1' -E_2)
   \longrightarrow \delta ( \sqrt{  {\bf p}_1^2 + M^2 }+\bar\nu - E_1')$ ;
by this way, the electromagnetic vertex of the
nuclear tensor  corresponds to that
of a free nucleon, evaluated at the same ${\bf
q}$, but at the transferred energy $\bar\nu$
instead of $\nu$ ,
which
 means that the nucleon hadronic tensor  has to be evaluated for
$p_N = p_N^{on}$ and $Q_N^2=\bar Q^2 = {\bf q}
^2 - \bar \nu^2 \neq Q^2$. By such a procedure
one obtains
\begin{eqnarray}
&& {d^3 \sigma \over  d \Omega ' d {\cal E}' }=
\int \overline \sigma_{eN} \ n^{D}(|{\bf p|}) d{\bf p}
\delta (\bar\nu +\sqrt{M^2+{\bf
p}^2}-\sqrt{M^2+({\bf p}+{\bf q})^2})=
\nonumber\\ && =(2\pi)
\int\limits_{|y|}^{p_{max}}
 \overline \sigma_{eN}\, \frac{E_{\bf p +q }}{|{\bf q}|}
 n^{D}(|{\bf p}|)|{\bf p}| d|{\bf p}|,
\label{pwia1}
\end{eqnarray}
where $\mbox{$\overline\sigma_{eN}$}$ is the
extrapolated electron -nucleon  cross section for
an off-mass shell nucleon \cite{forest},
$E_{\bf p+q}=E_1'=\sqrt{M^2+({\bf p}+{\bf q})^2}$,
and the limits of integration, which
are obtained from  the energy conservation provided
by the $\delta$- function, are as follows
\begin{eqnarray}
&&|{\bf p}|_{min} =  \frac{1}{2}\left | \left \{
(M_D+\nu)\sqrt{1-\frac{4m^2}{s}} -|{\bf q}|\right
\} \right | \equiv \left | y \right|
\label{ime093} \\ &&|{\bf p}|_{max} = \frac 12
\left \{ (M_D+\nu)\sqrt{1-\frac{4m^2}{s}} +|{\bf
q}|\right \} \equiv p_{max}, \label{ime094}
\end{eqnarray}
where $s$ denotes the Mandelstam variable for the
$\gamma^* D$ vertex
 \begin{equation}
s=(P_D+q)^2 =M_D(M_D+2\nu)-Q^2. \label{sman}
\end{equation}
and $y$ is the scaling variable according to \cite{ciofideut}
\begin{equation}
y =  \frac{1}{2}\left\{|{\bf q}|-
(M_D+\nu)\sqrt{1-\frac{4m^2}{s}}\right\}
\label{uai}
\end{equation}
When the value of
$|{\bf q}|$ becomes large enough, one has $p_{max}\sim \infty$ and the dependence of
 $\overline\sigma_{eN}$
 upon $|{\bf p}|$ becomes very weak. In such a case eq. (\ref{pwia1})
  can be cast in the following form \cite{ciofideut}
 \begin{equation}
 \frac{d\sigma}{d{\Omega}' d{\cal E}'}
=
 \left (s_{ep}+s_{en}\right )
 \frac{E_{y+|\vec q|} }{|{\bf q}|}\,
 (2\pi)\,
  \int\limits_{|y|}^{\infty}|{\bf  p}|  \, d|{\bf  p}|
 n^D(|{\bf  p}|),
 \label{nonrelappr}
 \end{equation}
where $s_{eN}$ and $E_{y +|\vec q| }$ represent
$\overline\sigma_{eN}$ and $E_{\vec p +\vec q }$,
calculated   at $|{\bf p}|=|{\bf p}|_{min}=|y|$, and can therefore
be taken out
of the integral. Such an approximation, which  has been
carefully investigated in ref. \cite{ciofideut}, turns out
 to be valid within few percents, provided
$Q^2 >0.5 \, GeV^2/c^2$.
 It is clear therefore, that  at large values of $|{\bf q}|$
  the following quantity
 (the {\it non relativistic scaling function})
\begin{eqnarray}
&& F(|{\bf q}|,y) \equiv \frac{|{\bf
q}|}{E_{y+|\vec q |}}\cdot \left(\frac{d\sigma}{
{d{\Omega}'d\cal E}'}\right)  / \left(s_{ep}
+s_{en}\right ) \label{scfunnon}
\end{eqnarray}
 will be directly related to the longitudinal momentum
 distribution
\begin{equation}
 F(|{\bf q}|,y)\,
\longrightarrow \,f(y)
=2\pi\int\limits_{|y|}^{\infty} |{\bf p}| d |{\bf
p}| n^D(|{\bf p}|), \label{add1}
\end{equation}
Thus  the
 condition for the occurrence of non relativistic
 $y$-scaling is that eq. (\ref{pwia1}) could be
cast in the form (\ref{nonrelappr}), which means
that: i) $ Q^2 > 0.5 \, GeV^2/c^2 $, in order to
make the replacement $\overline \sigma_{eN}\to \,
s_{eN}$ and $E_{\vec p+\vec q}\to\,E_{y+|\vec
q|}$ possible, and ii) $p_{max}=(|{\bf
q}|-|y|)\gg |y|$ (cf. (\ref{ime093}) and
(\ref{ime094})) in order to saturate the integral
of the momentum distribution,
$\int\limits_{|y|}^{p_{max}} |{\bf p}| d|{\bf p}|
n^D(|{\bf p}|)\to
 \int\limits_{|y|}^\infty |{\bf p}| d|{\bf p}| n^D(|{\bf p}|)$.
 Condition ii) obviously implies that the larger the value of
 $|y|$, the larger the value of $|{\bf q}|$ at which
 scaling will occur. The satisfaction of
 the inequalities $|{\bf q}|\gg 2|y|,\, x_{Bj} > 1 $
 leads, for any well-behaved $ n(|{\bf p}|)$, to the following
 conditions for the occurrence of non relativistic $y$-scaling:
 \begin{equation}
 2m/3\, \lesssim \nu\, < |{\bf q}|,\quad|{\bf q}|\gtrsim 2m.
 \label{conditions}
 \end{equation}
  Note, that
 the above conditions are very different from the conditions
 for Bjorken scaling $\nu \simeq |{\bf q}|$.

\section{The Final State Interaction}
\subsection {The  Schroedinger Approach}
In the calculation of the FSI, depicted in Fig. \ref{pict3}, it is more
convenient to perform calculations in the frame
where the interacting $np$-pair in the final state is at rest.
The phase-space factor can be written as follows
\begin{equation}
\frac{d {\bf p}_1'  d {\bf p}_2 }{E_1'E_2}
\delta^{(4)}(P_D+q-P_f) =\frac{ d {\bf P}_f d
{\bf p}_{rel}}{2E^{* 2}}\delta^{(3)}\left({\bf
q}-{\bf P}_f\right)\,
\delta\left(E^*-\frac{\sqrt{s}}{2}\right),
\label{delta}
\end{equation}
where ${\bf p}_{rel}$ is the relative momentum of
the $np$-pair which  is defined by the  Mandelstam variable
 $s=4\,({\bf p}_{rel}^2+M^2)$. For the
longitudinal current one has
\begin{eqnarray}
 G_E(Q^2) \exp(i{\bf q\,r}/2)= (4\pi)^2 G_E(Q^2)
\sum\limits_{\lambda,\mu} i^\lambda j_\lambda
(qr/2) {\rm Y_{\lambda \mu}^*(\hat q) Y_{\lambda
\mu}(\hat r)} \equiv (4\pi)^2
G_E(Q^2)\sum\limits_{\lambda,\mu}
 {\rm Y_{\lambda \mu}^*(\hat q)} {\rm \hat O_{\lambda \mu}} .
 \label{multipole}
 \end{eqnarray}
with   ${\rm \hat O_{\lambda \mu}}=  i^\lambda j_\lambda
(qr/2) Y_{\lambda
\mu}(\hat r)$, and the corresponding cross section is
\begin{eqnarray}
{d^3 \sigma^L \over  d \Omega' d {\cal E}'
}=\frac{4}{3} \frac{ M^2\, \sigma_{Mott} }{2\pi}
V_L\, G_E(Q^2)^2
\sum\limits_{J_f}\sum\limits_\lambda \left |
\langle J_D|| \hat O_\lambda(|{\bf q}|)
||p_{rel}; J_f L_f S_f \rangle \right |^2
 \frac{|{\bf p}_{rel}|}{\sqrt{s}}.
\label{wigner}
\end{eqnarray}
where  the radial
part of the two-nucleon wave function in the
continuum $| p_{rel};J_f L_f S_f \rangle  $
 has the following behaviour
\begin{equation}
u_{LS}^J(r)
\stackrel{r\to\infty}{\longrightarrow}\,\frac{1}{p_{rel}}\,
\sin \left(p_{rel}\, r- \frac{L\pi}{2}
+\delta_L\right). \label{asym}
\end{equation}
It can be seen that  equation (\ref{wigner})
differs  from the PWIA result (\ref{pwia1}).
 However, by using the
identity $\displaystyle\frac{1}{2|{\bf
q}|}\displaystyle\int\limits_{|y|}^{p_{max}}
\displaystyle\frac{| {\bf p}| d | {\bf
p}|}{E}=\frac{p_{rel}}{\sqrt{s}}$ one may  cast
the cross section in the following form
\begin{equation}
\frac{d\sigma^{L}}{d\Omega' d{{\cal E}'}}
=
 \left (s_{ep}+s_{en}\right )^{L}
 \frac{E_{y+|\vec q|} }{|{\bf q}|}\,
   \int\limits_{|y|}^{p_{max}}|{\bf  p}|  \, d|{\bf  p}|
 n_S^D(|{\bf  p}|, |\textbf{q}|, \nu),
 \label{xschr}
 \end{equation}
where, the following quantity has been introduced
\begin{equation}
 n_S^D(|{\bf  p}|, |\textbf{q}|, \nu)
=\frac{1}{4\pi}\,\frac{1}{3}\sum\limits_{J_f}\sum\limits_\lambda
\left | \langle J_D|| \hat O_\lambda(|{\bf q}|)
|| p_{rel};J_f L_f S_f \rangle \right |^2,
\label{newdist}
\end{equation}

\subsection {The Glauber Approach}
In the Glauber approach the exact two-nucleon continuum wave function
$|f>$ is approximated by its  eikonal  form.
Then the cross section can be written in the same
form as equation (\ref{pwia1}) with the deuteron momentum
distribution (\ref{deutwf}) replaced by the Glauber distorted
momentum distribution $n_G^D$~\cite{nikolaev},
\begin{equation}
n^D( {\bf p}) \to n_G^D( {\bf p}_m) =
\frac13\frac{1}{(2\pi)^3} \sum\limits_{{\cal
M}_D} \left | \int\, d  {\bf r} \Psi_{{1,\cal
M}_D}^*( {\bf r}) S( {\bf r}) \chi_f\,\exp (-i
{\bf p}_m {\bf r}) \right |^2, \label{ddistr}
\end{equation}
where
\begin{equation}
{\bf p}_m = {\bf q}-{\bf p}_1'  \label{missing}
\end{equation}
is the missing momentum,
 $\chi_f$  the spin wave
function of the final $np$-pair  and $S( {\bf r})$
 the $S$-matrix describing the  final state interaction
between the hit nucleon and the spectator, {\it {\it viz.}} (see Ref.
\cite{nikolaev})
\begin{equation}
S({\bf r}) = 1-\theta(z)\,\Gamma_{el}({\bf b}),
\label{sg}
\end{equation}
with the elastic profile function $\Gamma_{el}({\bf b})$
being
\begin{equation}
\Gamma_{el}({\bf b})=\frac{\sigma_{tot}(1-i\alpha)}{4\pi b_0^2}\,
\exp(-b^2/2b_0^2). \label{gama}
\end{equation}
In eqs. (\ref{sg}) and (\ref{gama}) ${\bf r} =
{\bf b} + z\, {\bf q}/|{\bf q}|$ defines the
longitudinal, $z$, and the  perpendicular, ${\bf b}$,
components of the relative coordinate {\bf r},
$\sigma_{tot}=\sigma_{el} + \sigma_{in}$, $\alpha$ is
 the ratio of the real to the
imaginary part of the forward elastic $pn$ scattering
amplitude, and, eventually,  the step function $\theta(z)$
originates from the Glauber's high energy
approximation, according to  which the struck nucleon
propagates along a straight-line trajectory and
can interact with the spectator only provided $z
> 0$. The following relations will be useful in what
follows
\begin{equation}
\sigma_{el} = \int \left | \Gamma_{el}({\bf
b})\right |^2d^2b=
\frac{\sigma_{tot}^2(1+\alpha^2)}{16\pi b_0^2}
\label{uno}
\end{equation}
\begin{equation}
 f_{el}({{\Delta}_{\perp}}) =
\frac{ik}{2\pi}\frac{\sigma_{tot}(1-i\alpha)}{4\pi
b_0^2} \int d^2 b\,  {\rm e}^{i{\bf \Delta b}}\, {\rm
e}^{-b^2/2b_0^2} = \frac{ik}{4\pi}
\sigma_{tot}(1-i\alpha) \, {\rm
e}^{-b_0^2\Delta_{\perp}^2/2}
\label{due}
\end{equation}
\begin{equation}
\frac{d\sigma_{el}}{d^2 \Delta} =\frac{1}{k^2} \left |
f_{el}({\bf \Delta})\right |^2=
\frac{\sigma_{tot}^2(1+\alpha^2)}{16\pi^2}
\exp(-b_0^2 \Delta_{\perp}^2)
\label{tre}
\end{equation}
where $\Delta$ is the transferred momentum in the $N-N$ collision, and
\begin{equation}
b_0^2=\frac{\sigma_{tot}^2(1+\alpha^2)}{16\pi\sigma_{el}}
\label{quattro}
\end{equation}
is the slope of the $q$-dependence of
the elastic proton-neutron cross section.
Assuming that  at high
relative energies of the $np$-pair the differences between the
absorbtion of  longitudinal (L) and
transverse (T) photons connected with the spin
dependence of FSI effects can be
disregarded, eq.(\ref{nonrelappr}) becomes
\begin{equation}
 \frac{d\sigma}{d\Omega' d{{\cal E}'}}
=
 \left (s_{ep}+s_{en}\right )
 \frac{E_{y+|\vec q|} }{|{\bf q}|}\,
 (2\pi)\,
   \int\limits_{|y|}^{p_{max}}|{\bf  p}_m|  \, d|{\bf  p}_m|
 n_G^D(|{\bf  p}_m|, \cos\theta_{{\bf qp}_m}).
 \label{glauberfy}
 \end{equation}
It should be stressed, first, that in absence of any FSI,
the distorted momentum distribution  $n_G^D({\bf
p}_m)$ reduces to the undistorted momentum distribution
 $n^D({\bf p})$  (${\bf p}_m =-{\bf p}_1$) and, secondly,  that unlike   $n^D({\bf
p})$,   $n_G^D({\bf
p}_m)$ depends also upon the orientation
of ${\bf  p}_m$ with respect to the momentum transfer
 ${\bf  q}$, with  the angle $\theta_{{\bf qp}_m}$ being fixed by the
 energy conserving $\delta$-function, namely
 $\cos \theta_{{\bf qp}_m}=[(2(M_D+\nu)\sqrt{|\textbf{p}_m|^2+M^2}
 -s)]/(2|\textbf{q}||\textbf{p}_m|)$;
 thus
 $n_G^D({\bf  p}_m)$
depends implicitly on the kinematics of the process, and
the values of  $y$ and  $|{\bf  q}|$
fix the value of the total energy
(\ref{sman}) of the final $np$ pair, i.e. the
relative energy of the nucleons in the final
states. Consequently, the quantities
$\sigma_{tot}$, $\alpha$ and $b_0$ in
(\ref{gama}) also depend upon the kinematics of
the process. In this sense, the distorted
momentum distribution $n_G^D({\bf  p}_m)$ implicitly
depends upon $|{\bf  q}|$ and $y$ as well.

\subsection{The longitudinal sum rule}
Let us now briefly discuss the charge
conservation sum rule in the quasi-elastic
processes. The longitudinal part of the
hadronic current is the charge density of the
target and the longitudinal cross section may be
written in the form
\begin{eqnarray}
&& {d^3 \sigma^L \over  d \Omega'  d {\cal E}'
}=\int \frac{V_L}{3} \sum\limits_{{\cal
M}_D,J_f,{\cal M}_f} |\langle P_D,{\cal M}_D|\hat
J_L^D(Q^2)|P_f,{\cal M}_f\rangle|^2 \left [
\frac{ d {\bf p}_2  }{(2\pi)^3} \delta (\nu
+E_i-E_f) \right ]. \label{long}
\end{eqnarray}
Integrating over the energy loss $\nu$, summing  over the final states and,
 disregarding, for ease of presentation,   the neutron
form factor $G_E^n$, the longitudinal sum rule
can be obtained (see for details
ref.\cite{ciofiprogress})
\begin{equation}
{\cal S}= \int\, \left ( G_E^2(Q^2)\, V_L\right
)^{-1} \frac{d^3 \sigma^L}{ d \Omega  d {\cal E}'
}\, d\nu = \frac{1}{3} \sum\limits_{{\cal M}_D}
\int \left | \int\, \Psi_{1,\cal M}({\bf r})\, \exp (i{\bf
pr})) d {\bf r}\right |^2 \frac{ d {\bf p}
}{(2\pi)^3} = 1. \label{sr}
\end{equation}
Note  that the sum  over the final
states contains also the contribution from
elastic scattering, so that in order to obtain
the longitudinal sum rule corresponding to the inelastic
scattering the
elastic part, $F^2_p(Q^2)=|\langle D| \exp(i{\bf
qr}_p)| D\rangle |^2$, has to be subtracted
 from eq. (\ref{sr}), obtaining
\begin{equation}
{\cal S}_{in}=\lim\limits_{Q^2\to\infty} \left (
{\cal S} -F^2_p(Q^2)\right )\,
\longrightarrow\,\,\, 1. \label{sr1}
\end{equation}

 The longitudinal sum rule (\ref{sr1})
is fulfilled exactly within the PWIA,  as well as
when the  Schroedinger approach is used to include the FSI;
if the latter are considered within  the Glauber
approach, as described in the previous paragraph, the sum rule is not satisfied.
 As a matter of fact  by using eqs. (\ref{uno})-(\ref{quattro}) and
introducing the inelastic profile function $\Gamma_{inel}({\bf b})$
through the unitarity relation
\begin{eqnarray}
2 Re \,\Gamma_{el}({\bf b}) = \Gamma_{el}({\bf b})+\Gamma_{inel}({\bf b}),
\label{gl}
\end{eqnarray}
one obtains
\begin{equation}
{\cal S}_{in}=\int d{\bf r} \left |  \Psi_{1,\cal M}({\bf
r}) \right |^2 \left ( 1-\theta(z)\left |
\Gamma_{inel}({\bf b})\right |^2\right ),
\label{glsr}
\end{equation}
which shows  that  if  the
inelastic channels are absent, the longitudinal sum rule
(\ref{sr})
is fulfilled, whereas in the presence of open
inelastic channels one has $ {\cal S}_{in} < 1$,
i.e. the
incident nucleon flux is partially absorbed by
inelastic processes.

\section{Results of calculations}
\subsection{ The Schroedinger approach.}
 The
calculation of the cross section and the scaling
function by eqs. (\ref{xschr}) and
(\ref{scfunnon}), requires the knowledge of  the wave
functions $| p_{rel};J_f L_f S_f \rangle$ of the
final $np$-pair, which are  solutions of the
Schroedinger equation in the continuum with a
given nucleon-nucleon potential. It is well known
that the non relativistic
deuteron momentum distributions calculated with
different realistic potentials, {\it viz.} the
 Bonn \cite{bonn}, Paris
\cite{paris}
and Reid \cite{Reid} ones, exhibit   rather different
behaviours  at moderate and large momenta. It has
also  been shown that relativistic calculations
of the deuteron momentum distribution within the
Bethe-Salpeter formalism,  yield  results which are very
close to
those obtained with the Reid Soft Core (RSC)
potential (see ref.~\cite{ciofiBS}). Therefore we
have used
 the  RSC potential to solve the
Schroedinger equation for the $| p_{rel};J_f L_f
S_f \rangle$ states, taking into account  all
partial waves with $J_f < 3$. For higher values of  $J_f$
the PWIA has been  adopted. For the
sake of  comparison with the experimental data
we have also assumed that the effects of the
 FSI on  the longitudinal and
transverse parts of the cross section is the
same, and is governed by the quantity
(\ref{newdist}). In the Schroedinger approach,
FSI arise from the
elastic rescattering of the two nucleons in the
final states.
 The threshold
for inelastic channels  corresponds
to a value  of the total energy of the $np$-pair
$\sqrt{s} \gtrsim 2\,  GeV$, or equivalently,
$p_{lab} \gtrsim 0.8\, GeV/c$, where $p_{lab}$ is
the laboratory momentum of the struck nucleon
(i.e. with the spectator at rest),
 corresponding to a total energy
$\sqrt{s} =
\sqrt{2M^2+2M\sqrt{p_{lab}^2+M^2}}$. Experimentally \cite{baldini}, the
inelastic channel contribution starts to be relevant at
$p_{lab} \simeq 1.2\, GeV/c$.
 The inclusive $D(e,e')X$ cross section corresponding to
 electron beam energy ${\cal E} = 9.761 GeV$
 and scattering angle
 $\theta = 10^0$ is shown in fig. \ref{xshr}. The dotted
  line corresponds
to the PWIA and  the solid curve
is the result which includes the FSI.   It can be
 seen that in the range  $0.8\, GeV < \nu < 1.2\, GeV$,
 FSI increases the cross section  and substantially
improve the description of the data; on the contrary,  near the
quasi-elastic peak  FSI decrease the cross section,
as it should be, since in agreement with  the sum rule
(\ref{sr1}) the integral over $\nu$ must be
conserved. Our results fully agree
with those obtained in Ref. \cite{arenhovel}.
In  the kinematics we have considered
the variation of  $\nu$, from  threshold
to the quasi-elastic peak, corresponds to a variation
 of $p_{lab}$ in the range  $0.6\,
GeV/c\, <\, p_{lab}\, < 2\,GeV/c $ (cf. the upper scale in
 fig. \ref{xshr} and Table \ref{tablitza})
 where the elastic nucleon-nucleon
scattering still dominates. Note, that in this
case the corresponding values of $y$ and $|{\bf
q}|$ change in the range  $-500 \, MeV/c\, < \,y\, \lesssim 0 \,$
and $3.3 \, GeV^2/c^2 \,< \,|{\bf q}|^2\,< 4.3\, GeV^2/c^2 $
 respectively. Let us now keep $y$ fixed
and vary the values of  $|{\bf q}|$, i.e.
 check the effects of FSI on the scaling function  $F(|{\bf q}|,y)$,
 defined by
eqs. (\ref{scfunnon}). The results are presented in fig.
\ref{scaleschr}.
The dotted line is the scaling function within the
PWIA, and  the solid curve includes the effects
of FSI.  On the top horizontal axes the
corresponding value of  $p_{lab}$ is also
shown. At low  values of
$|{\bf q}|$ the effects of FSI are very large and no
scaling behaviour can be observed.
With increasing $y$, the scaling violation near
the threshold values of $|{\bf q}|$,
 increases.  This is due to the fact that a larger value of $y$
results in a  lower value of  $p_{lab}$, in correspondence of which
 the elastic
cross section is much higher ~\cite{baldini}.
 FSI decreases with  $|{\bf q}|$,
and at values corresponding to   $p_{lab} \gtrsim
1\,GeV/c$ the function $F(|{\bf q}|,y)$ exhibits  a
scaling behaviour. It should be stressed, that
values of
 $p_{lab}\sim 1 \,GeV/c$ are
 still in the kinematics  region where  the Schroedinger
 approach can be  applied.
 At asymptotic values, $|{\bf q}|\to\infty$,
 the total energy of the $np$-pair
$\sqrt{s} \to \infty$, consequently the phase
shifts $\delta_L$ in eq. (\ref{asym}) vanish and
the final states $| p_{rel};J_f L_f S_f \rangle$
become just the partial decomposition of plane
waves, so that the Schroedinger approach and the
PWIA coincide.

\subsection{The Glauber approach}
The
inclusive  cross section  calculated within the Glauber approach, using
the  RSC and  Bonn potentials,  is
presented in fig.~\ref{xgl}. It can be seen that: i)the two potentials
 give very different
results at low values of $\nu$, ii) Glauber FSI appreciably depend
upon the potential model.
Our analysis shows that
such a potential dependence in the
kinematical region  at low $\nu$, can be explained by considering that
the corrections to the  deuteron $S$ and $D$-waves generated by the FSI
are opposite in sign. In the Glauber approach FSI are entirely driven
by  the distorted momentum distribution $n_G^D$; let us
therefore   discuss the properties of the latter
 within the kinematical conditions
relevant to $y$-scaling (for a detailed analysis
of $n_G^D$ at asymptotic energies see
ref.\cite{nikolaev}). It turns out that
$n_G^D$ depends upon $p_{lab}$, which is a function of  $y$ and $|{\bf q}|$.
More explicitly,
the $p_{lab}$-dependence  of $n_G^D$  arises   from the $p_{lab}$-dependence
 of the parameters $\alpha$ and $b_0$,  appearing in
  the profile function $\Gamma_{el}({\bf b})$ ((\ref{gama}));
  such a dependence is shown in Figure \ref{alfab0}.
 It can be  seen that when the energy is  high enough,
 ($p_{lab}\gtrsim 1.5-2\, GeV/c$), the parameters $\alpha$ and $b_0$
 become almost
constant and, consequently, the distorted momentum
distribution $n_G^D$ becomes independent
of the  kinematics of the process. In the region
$p_{lab} < 2\, GeV/c$,  the parameters  $\alpha$
and $b_0$ exhibit a strong $p_{lab}$ dependence, and so does the
momentum distribution $n_G^D$. The $|{\bf q}|$ dependence
of $n_G^D$ calculated  at $| {\bf p}_m|
= |y|$ and $\theta_m =0$
is shown in  fig.~\ref{distotq}. It turns out that:
 i)the undistorted momentum distributions $n^D$ at large values
 of $y$ strongly depend upon the potential model; ii) $n_G^D$ exhibits
 a strong $|{\bf q}|$ behaviour at  low values of ${\bf q}$; iii)
 at high values of  $|{\bf q}|$ ( which correspond to high
values of $\sqrt{s}$ and $p_{lab}$)
  the distorted momentum
distribution    $n_G^D$ scales to a quantity  which, at large negative values of
$y$, may  differ
from the undistorted momentum distributions  $n^D(|y|)$ (the straight lines in fig.
\ref{distotq}),
at variance with the Schroedinger result, which predicts   $n_S^D \simeq n^D(|y|)$
 at high values of $|{\bf q}|$;
 iv) at high values of $y$ the potential model dependence of
 $n^D(|y|)$.
The explanation of points i) and ii)  is clear:
at low values of  $|{\bf q}|$ the Glauber FSI is driven by the
elastic cross section, which strongly decreases with $|{\bf q}|$;
with increasing  $|{\bf q}|$, $p_{lab}$ reaches the inelastic threshold value
( $p_{lab} \simeq 0.8 GeV/c $) and  the total  cross section scales to its
asymptotic value
$\sigma_{tot}\sim\, 44 \, mb$ ($\alpha=-0.4$, $b_0=0.5 \, fm$), and so does   $n_G^D$.
The possible reasons for  the differences between the asymptotic   $n_G^D$ and
  $n^D(y)$ (point iii)) will be  briefly discussed later on.
The comparison between the Schroedinger and the Glauber
approaches for the scaling  function $F(|{\bf q}|,y)$ is shown
in fig.~\ref{fyglaub}. It can be seen that, for large values of $y$,
and   below the pion production
threshold ($p_{lab} \simeq 0.8 GeV/c$), which is the region of existing
 experimental data,
 the Schroedinger approach
provides a satisfactory description of the experimental scaling function
$F(|{\bf q}|,y)$
, unlike
the Glauber approach, which overestimate the data  at low $|{\bf q}|$
and underestimate them at high  $|{\bf q}|$. The difference between
the Schroedinger and Glauber results is strongly reduced at low values
of $y$ ($x_{Bj} \simeq 1$), where, being the target nucleon almost
free,
the small-scattering-angle requirement necessary for the validity of
the Glauber approximation, is probably better fulfilled.

A common approximation, adopted by various authors  in the Glauber type
 calculation of the FSI,
is to consider that at $Q^2 \simeq 1 GeV^2$ the asymptotic
$\sigma_{tot}\sim\, 44 \, mb$ should be used.
The validity of such an approximation is illustrated in
 Figure~\ref{fyconst}, where the dashed line represents
  the
results obtained using the  asymptotic $n-p$ cross section,
the full  lines the  results with the  quantities
$\alpha$, $b_0$ and $\sigma_{tot\,( el)}$  which properly include the dependence upon
the  relative momentum $p_{lab}$, and the dotted   line the   PWIA.

\section{Summary and Conclusions}
The aim of this paper was to address the longstanding
problem of the evaluation of  FSI effects in
  inclusive  processes
$A(e,e')X$,  which have been described, to date,
by  various approximate
approaches. To this end,
we have considered the electro-disintegration of the deuteron,
and have performed exact  calculations within two different approaches
 to treat the final state, viz:
i) the Schroedinger approach, in which, given a realistic
two-nucleon interaction, the Schroedinger equation is solved
to generate bound and continuum two -nucleon states, with the latter
describing elastic $n-p$ rescattering, and ii) the Glauber
high energy approximation, paying, in this case, particular
attention   to a correct treatment of the
 kinematics. Our aim was to understand the limits of validity
 of the two approaches, and to pin down the main features
 of the FSI mechanism, having also in mind a better understanding
 of these effects in complex nuclei,  where calculations cannot
 be performed exactly. From the calculations we have exhibited,
 the following remarks are in order:

 1) the existing experimental data on the $D(e,e')X$ process
 at $x_{Bj}>1$ (negative values of $y$) are, to a  large extent,
 limited to a kinematical range where the invariant
  mass
 of
 the final hadronic state  $\sqrt s$  is below the inelastic
 channel threshold  ${s}  \lesssim  4  GeV^2$ (or $p_{lab}  \lesssim
 0.8 GeV$) (cf  fig.~\ref{fyglaub} and Table \ref{tablitza});
 therefore, in spite of the large value of $Q^2$ involved,
the two nucleons in the continuum mostly undergo elastic
 scattering, so that   the   Schroedinger approach should represent
 the correct description of the process and, as a matter of fact, the
 calculations describe the experimental data rather well.

 2) The Glauber results overestimate the Schroedinger results
  at low values of
 $|{\bf q}|$, and underestimate them at high values
 of  $|{\bf q}|$. The reason for such a disagreement between the
 two approaches, which is particularly relevant
 at large values of $x_{Bj}>1$ (large, negative
 values of $y$), has to be  ascribed
 to the fact that at $x_{Bj}>1$,
 the direction of the ejected nucleon  sizably differs from the
 direction of the momentum transfer.

 3) At values of  ${s}  \gtrsim  4  GeV^2$ (or $p_{lab}  \gtrsim
 1 GeV$), i.e. above the pion production  threshold, both the
 Schroedinger
  and the Glauber approaches might become inadequate, for the
  propagation of nucleon excited states (inelastic rescattering)
   have to be explicitly taken
  into account. Calculations of this type, within the approach
  proposed in Ref. \cite{braun}, are in progress and will be reported
  elsewhere.

 \acknowledgements
\vskip.15in
This work was partially supported by the Ministero dell'Universit\`a e della
Ricerca Scientifica e Tecnologica (MURST) through the funds COFIN99.
 Discussions with  M. Braun,  S. Dorkin and B. Kopeliovitch are gratefully acknowledged.
  L.P.K. thanks  INFN,  Sezione di Perugia, for
  warm hospitality and financial support.

\newpage
\begin{figure} 
\caption{ The One Photon Exchange diagram for the
   $D(e,e' )X$ process.}
\label{pict1}
\end{figure}

\begin{figure} 
\caption{ The PWIA diagram for the
quasi elastic  $D(e,e' )X$ process. }
\label{pict2}
\end{figure}

\begin{figure} 
\caption{ The FSI diagram for the
quasi elastic  $D(e,e' )X$ process. }
\label{pict3}
\end{figure}

\begin{figure} 
\caption{ The inclusive cross section $D(e,e' )X$
versus the energy transfer $\nu$ and  the laboratory momentum of
the struck nucleon in the final state  $p_{lab}$
(note that the inelastic threshold corresponds to  $p_{lab}\simeq 0.8\,
Gev/c$). Dotted line:  PWIA calculation;
full
line: effects of the FSI, calculated within the
Schroedinger approach~(\protect\ref{xschr}). The
experimental data,  from Ref.  \protect\cite{Bosted},
correspond to   electron initial energy
${\cal E}=9.761\, GeV$ and  scattering angle
$\theta=10^o$\,.}
\label{xshr}
\end{figure}

\begin{figure} 
\caption{  The scaling function $F(|{\bf q}|,y)$,
eq. (\protect\ref{scfunnon}),
for various values
of $y$
vs. the three-momentum
transfer $|{\bf q}|$ and  $p_{lab}$.
Dotted line: PWIA; solid line:  FSI
within the  Schroedinger approach. In this and the following Figures,
the values of the other relevant kinematical variables, e.g.
$\nu =-M_D+\sqrt{y^2+M^2}+\sqrt{(y+|{\bf q}|)^2+M^2}$,
$\, \, Q^2=|{\bf q}|^2-\nu^2$,  $x_{Bj}=Q^2/2M\nu$,
$\sigma_{el}$ and $\sigma_{tot}$, can be found in Table~\ref{tablitza}.
 The experimental scaling function is from Ref. \protect\cite{ciofideut} }
\label{scaleschr}
\end{figure}

\begin{figure} 
\caption{
The inclusive cross section $D(e,e' )X$ calculated within the Glauber
approach.
The deuteron wave function corresponds to the
RSC and Bonn potentials. The
experimental data are the same as in fig.\ref{xshr}}
 \label{xgl}
\end{figure}

\begin{figure} 
\caption{The ratio $\alpha$ between  the imaginary to the
real part of the forward elastic amplitude for
 $np$-scattering, and the
parameter $b_0$, eq. (\ref{quattro}),
 used in the parameterization of the
profile function  (\ref{gama}). The
experimental data for $\alpha$ are taken from
\protect\cite{said}} \label{alfab0}
\end{figure}

\begin{figure} 
\caption{The dependence of the distorted momentum
distribution $n_G^D$, eq. (\ref{ddistr}),
 upon $|{\bf q}|$ for various values of $y$ and fixed
  values of
 $\theta_m=0^o$ and $|{\bf p}_m|=|y|$. The solid line corresponds to the
  Reid potential and the dashed line to the Bonn potential.
 The  dotted (dot-dashed) line represents
  the corresponding RSC (Bonn)  undistorted momentum distributions
 $n(|y|)$ ( eq. (\ref{deutwf})).}
\label{distotq}
\end{figure}

\begin{figure} 
\caption{ The scaling function $F(|{\bf q}|,y)$
vs $|{\bf q}|$ and $p_{lab}$,  for various values of $y$,
corresponding to  the Glauber (full)
 and the
Schroedinger(dashed) approaches, respectively.
 The dotted line  represents
 the PWIA. The experimental data are the same as in fig.\ref{scaleschr}.
All curves correspond to the RSC potential.}
\label{fyglaub}
\end{figure}

\begin{figure} 
\caption{ The scaling function $F(|{\bf q}|,y)$
vs $|{\bf q}|$ for various values of $y$  and $p_{lab}$. The
full line was obtained using in the  Glauber
approach the correct dependence upon  $p_{lab}$
of the quantities $\alpha$, $b_0$,
$\sigma_{tot}$ and
$\sigma_{el}$, whereas
the dashed line has been  obtained with the
 asymptotic values
 $\alpha=-0.4$, $b_0=0.5\,fm$ and
$\sigma_{tot}=44.2\,mb$.
 The dotted  line
represents  the PWIA. All curves correspond to the RSC potential.}
 \label{fyconst}
\end{figure}

\newpage
\epsfxsize 12cm         
\centerline{ \epsfbox{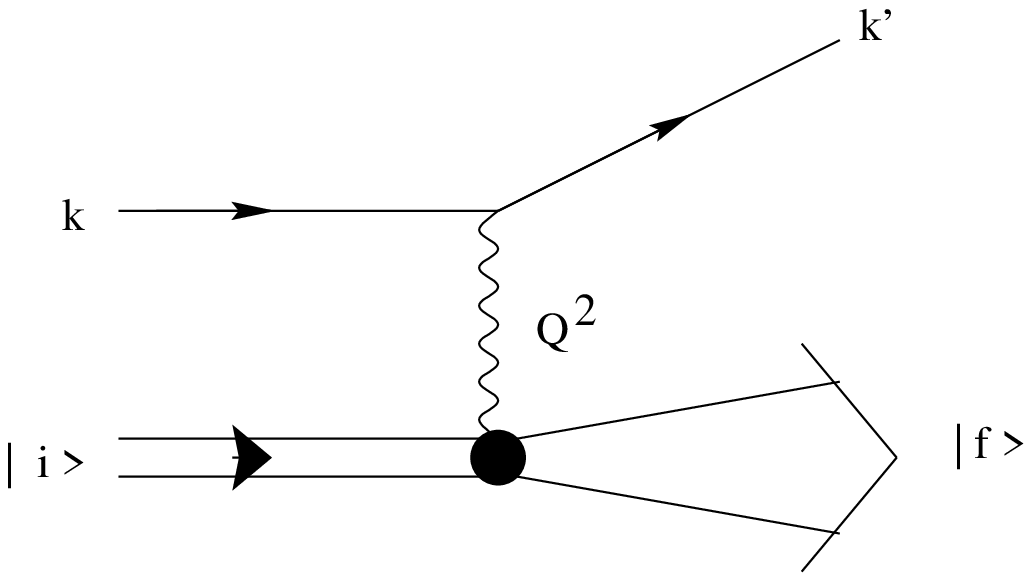}}
 \vskip 2.2cm
\hspace{3cm}
Fig.~\ref{pict1}. C. Ciofi degli Atti....On the FSI effects....

\epsfxsize 8cm         
\centerline{ \epsfbox{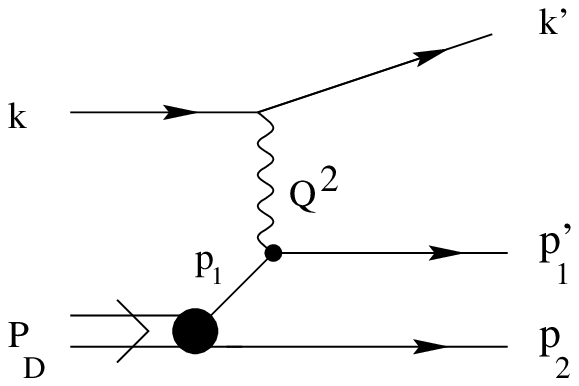}}
 \vskip 2.2cm
\hspace{3cm}
Fig.~\ref{pict2}. C. Ciofi degli Atti....On the FSI effects....

\epsfxsize 8cm         
\centerline{ \epsfbox{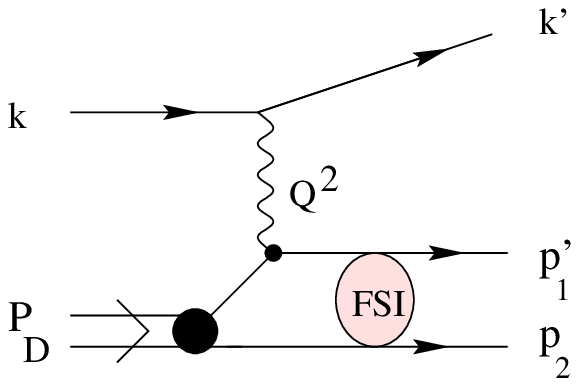}}
 \vskip 2.2cm
\hspace{3cm}
Fig.~\ref{pict3}. C. Ciofi degli Atti....On the FSI effects....

\newpage
\epsfxsize 15cm         
\centerline{ \epsfbox{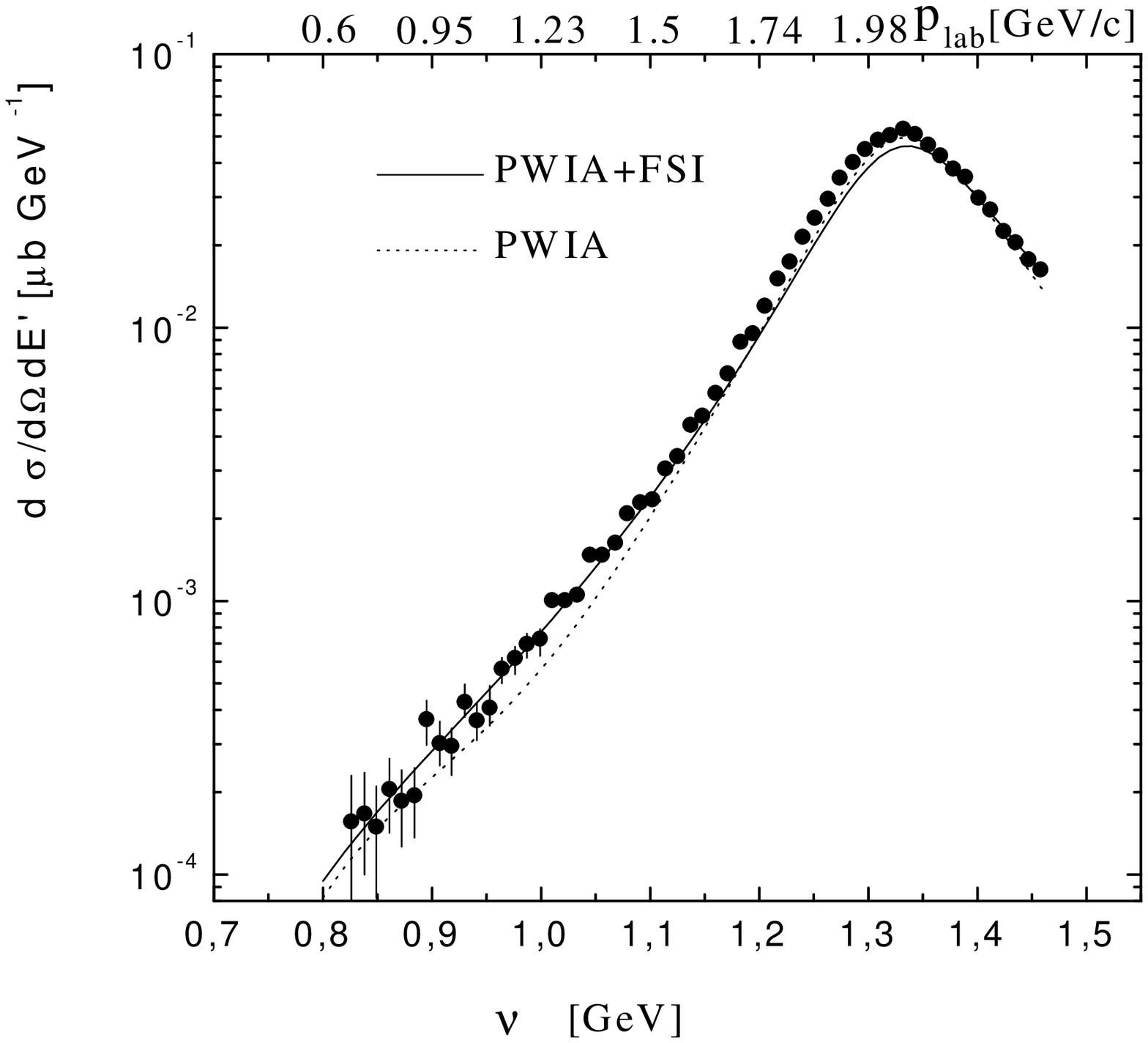}}
 \vskip 2.2cm
\hspace{3cm} \vfill
Fig.~\ref{xshr}. C. Ciofi degli Atti....FSI
effects....
\newpage

\epsfxsize 14cm  
\centerline{ \epsfbox{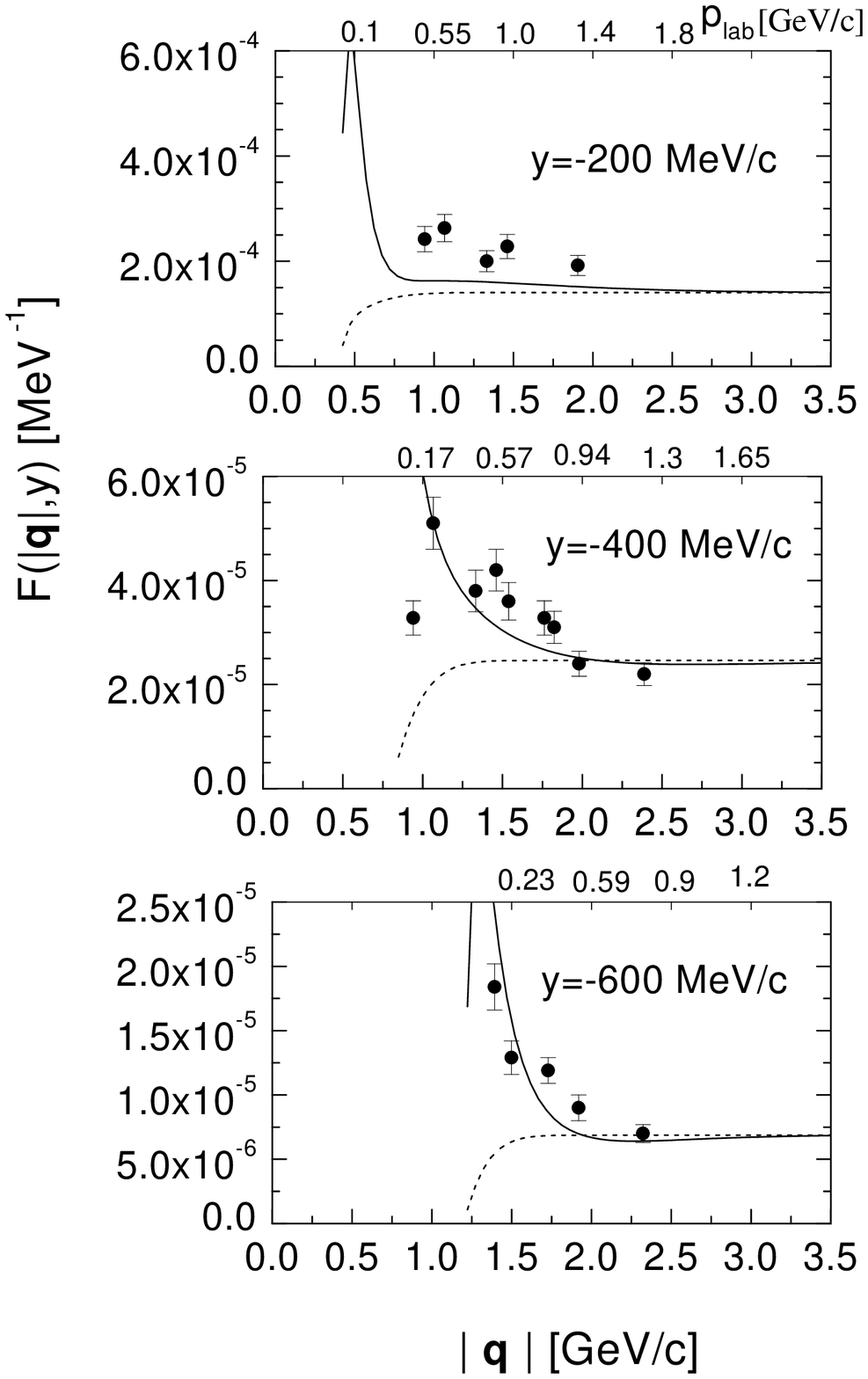}}
\vfill

Fig.~\ref{scaleschr}. C. Ciofi degli Atti....FSI
effects..

\newpage
\epsfxsize 16cm  
\centerline{\epsfbox{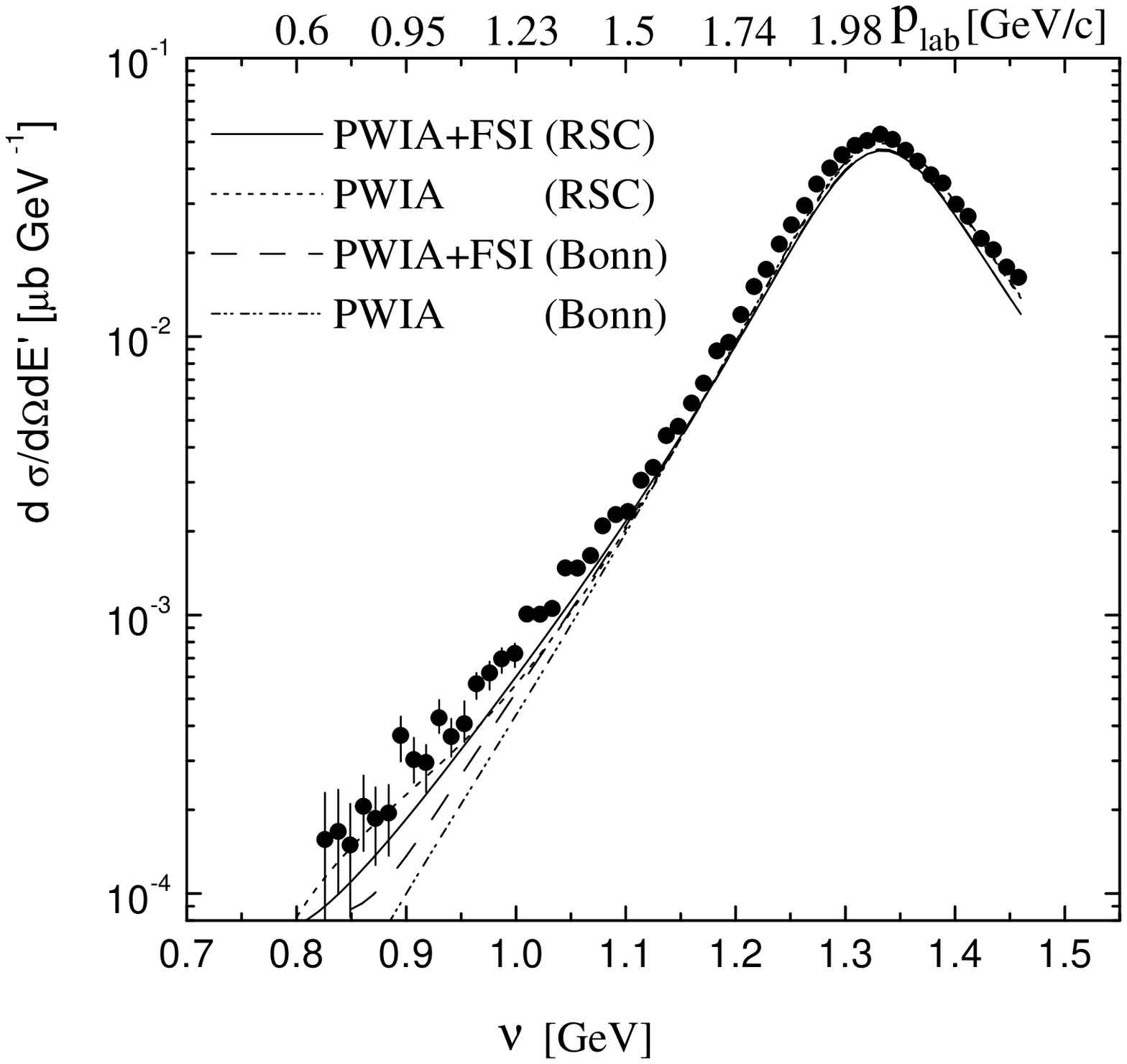}}
\vfill

Fig.~\ref{xgl}. C. Ciofi degli Atti....FSI
effects..
\newpage
\epsfxsize 15cm          
\centerline{ \epsfbox{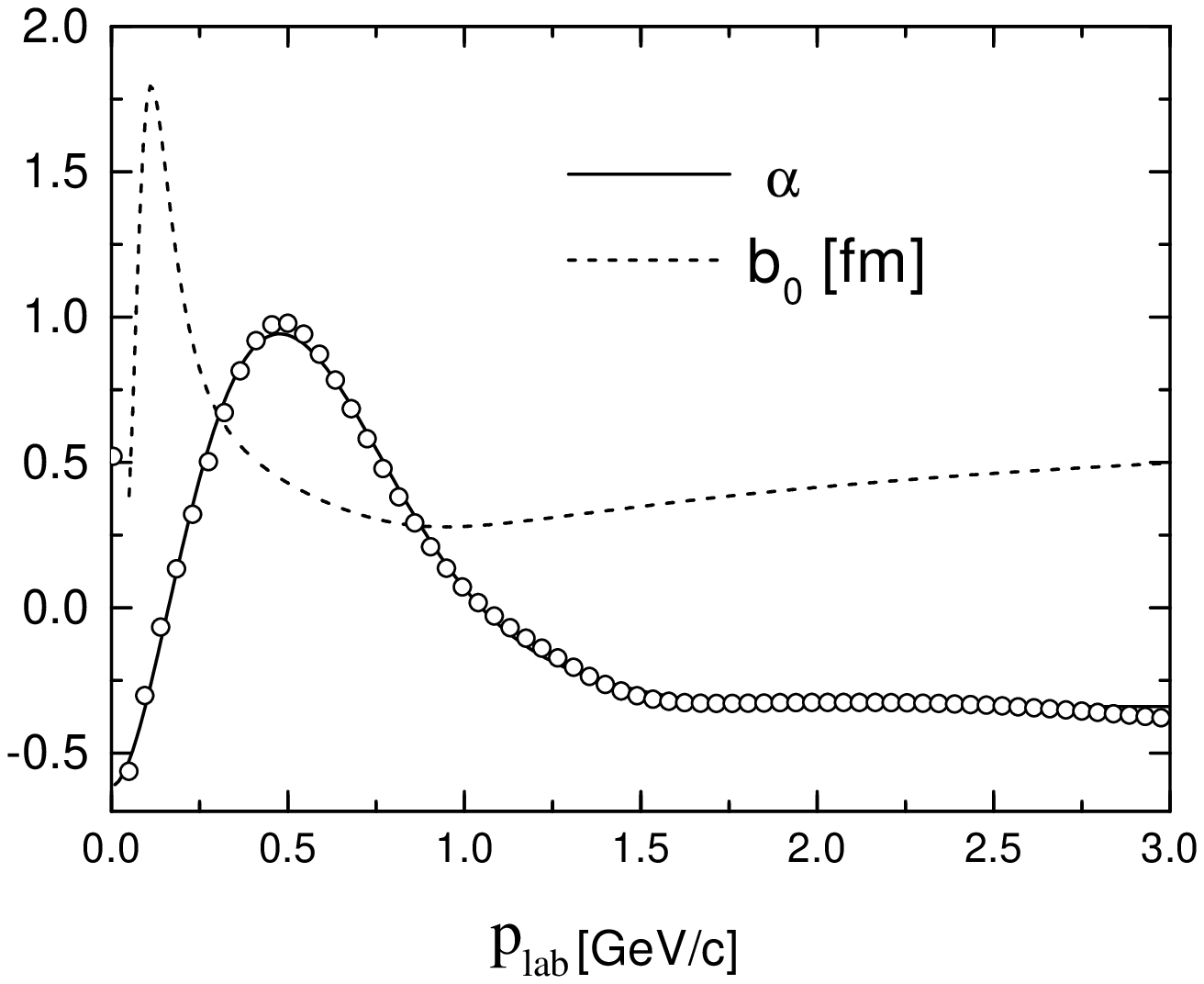}}
 \vskip 2.2cm
\hspace{3cm} \vfill

Fig.~\ref{alfab0}. C. Ciofi degli Atti....FSI
effects..

\newpage
\epsfxsize 15cm 
\centerline{ \epsfbox{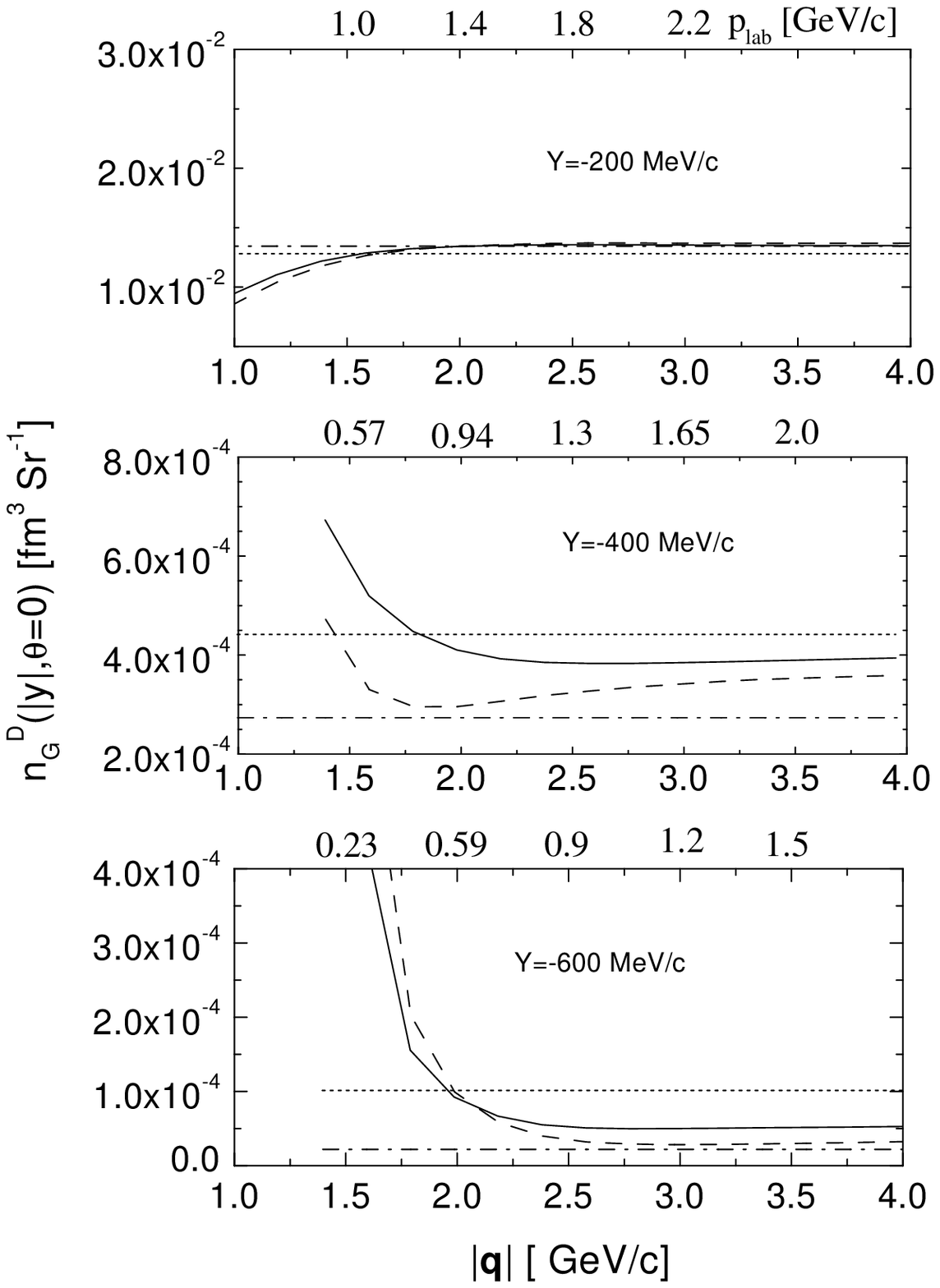}}
 \vskip 0.2cm
\hspace{3cm} \vfill

Fig.~\ref{distotq}. C. Ciofi degli Atti....FSI
effects..


\newpage
\epsfxsize 13cm        
\centerline{ \epsfbox{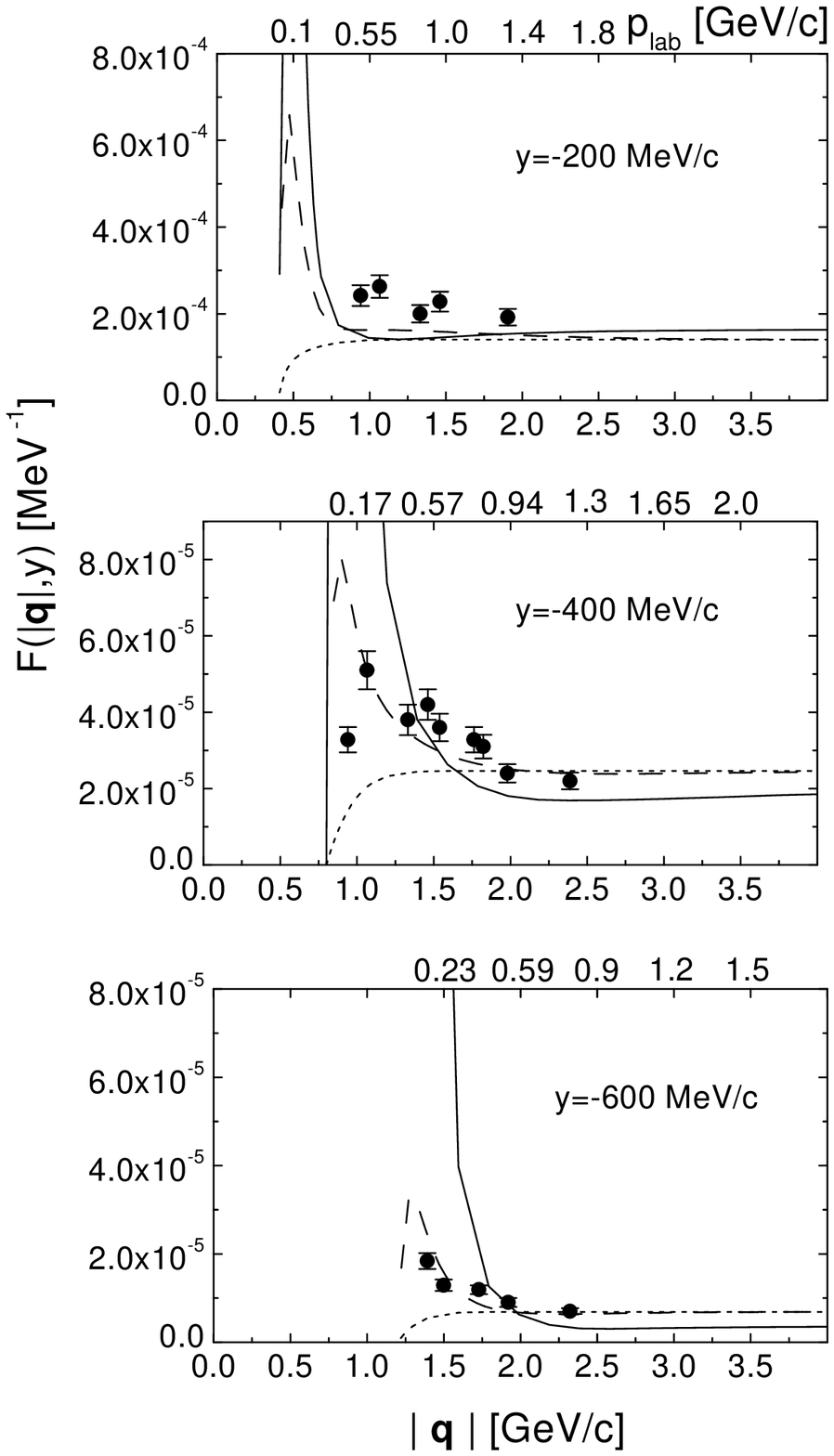}} \vfill
Fig.~\ref{fyglaub}. C. Ciofi degli Atti....FSI
effects..

\newpage
\epsfxsize 12cm        
\centerline{ \epsfbox{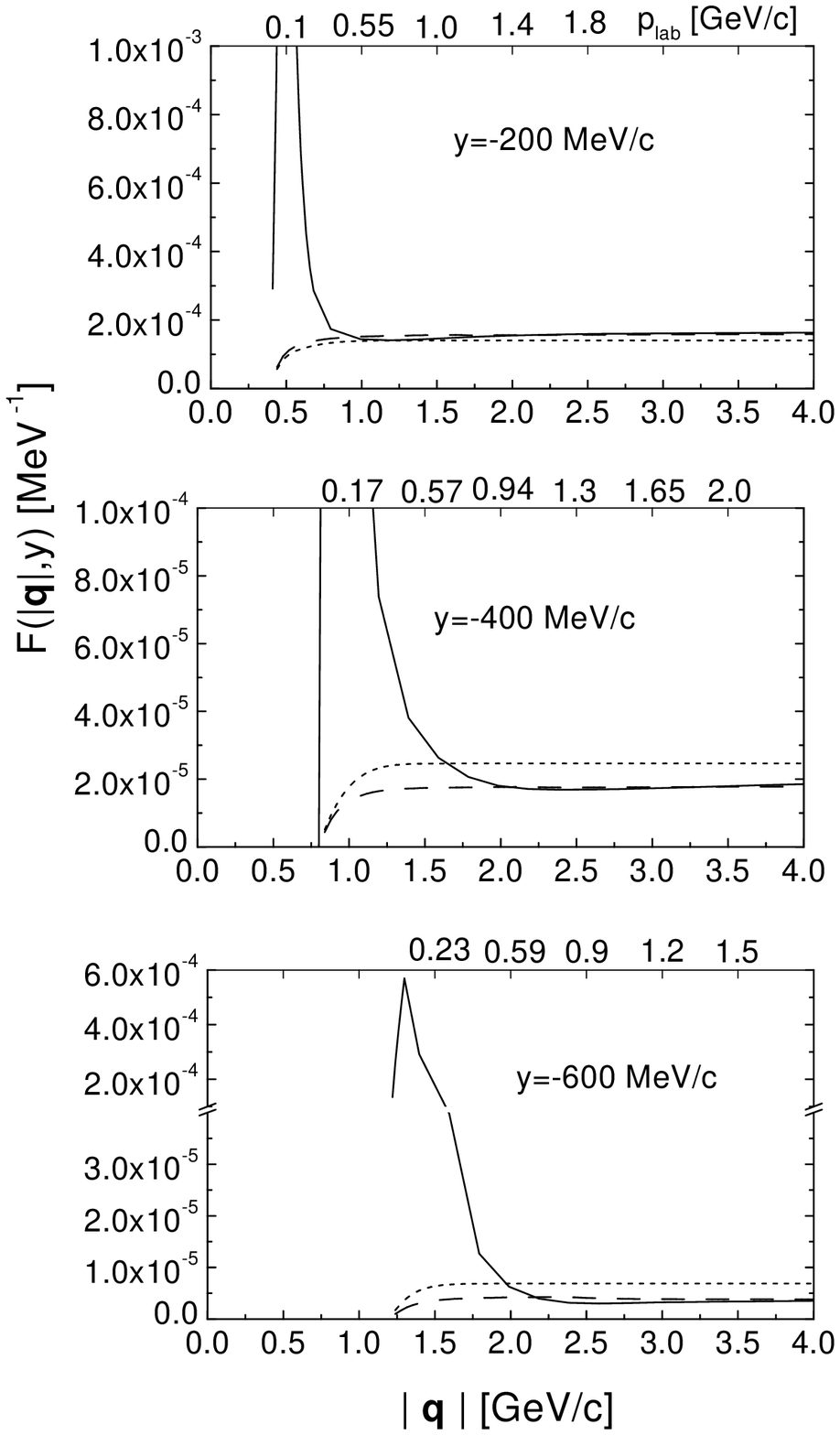}} \vfill
Fig.~\ref{fyconst}. C. Ciofi degli Atti....FSI
effects..

\begin{table*}[t]
\centerline{y=-200 MeV/c}
\begin{tabular}{|cccccccc|}
$|{\bf q}|$ & $\nu$ & $Q^2 $ & $x_{Bj}$  & $p_{lab} $ & $ s$ & $\sigma_{el}$ & $\sigma_{tot}$ \\
$ GeV/c$ & $ GeV$ & $GeV^2/c^2$ &    & $ GeV/c$ & $  GeV^2$ & $ mb$ & $mb $\\
  \hline\hline
   .50 &      .07&       .25&      1.86&       .10&      3.53&   1744.52  &    1744.52 \\
   .85 &      .23&       .67&      1.58&       .42&      3.69&     94.85  &      94.85 \\
  1.20 &      .46&      1.23&      1.44&       .73&      3.99&     45.58  &      45.58 \\
  1.55 &      .73&      1.87&      1.37&      1.03&      4.37&     31.96  &      35.71 \\
  1.90 &     1.03&      2.56&      1.33&      1.32&      4.81&     25.85  &      35.78 \\
  2.25 &     1.34&      3.27&      1.30&      1.62&      5.27&     22.44  &      37.06 \\
  2.60 &     1.66&      4.00&      1.28&      1.90&      5.74&     20.29  &      38.57  \\
  2.95 &     1.99&      4.74&      1.27&      2.19&      6.24&     18.82  &      39.99 \\
  3.30 &     2.32&      5.49&      1.26&      2.48&      6.73&     17.76  &      41.19  \\
  3.65 &     2.66&      6.25&      1.25&      2.77&      7.24&     16.95  &      42.17  \\
  4.00 &     3.00&      7.00&      1.24&      3.05&      7.75&     16.32  &      42.92
  \\
\end{tabular}
\end{table*}
\begin{table*}
\centerline{y=-400 MeV/c}
\begin{tabular}{|cccccccc|}
$|{\bf q}|$ & $\nu$ & $Q^2 $ & $x_{Bj}$  & $p_{lab} $ & $ s$ & $\sigma_{el}$ & $\sigma_{tot}$\\
$ GeV/c$ & $ GeV$ & $GeV^2/c^2$ &    & $ GeV/c$ & $  GeV^2$ & $ mb$ & $mb$\\
\hline\hline
    .90  &       .21 &        .77 &       1.96 &        .09 &       3.53 &2057.63      &  2057.63\\
   1.25  &       .41 &       1.39 &       1.80 &        .38 &       3.66 & 109.40      &   109.40\\
   1.60  &       .67 &       2.11 &       1.68 &        .66 &       3.91 &  51.82      &    51.82\\
   1.95  &       .96 &       2.89 &       1.61 &        .91 &       4.22 &  35.89      &    36.61\\
   2.30  &      1.26 &       3.69 &       1.56 &       1.16 &       4.56 &  28.72      &    35.47\\
   2.65  &      1.58 &       4.52 &       1.52 &       1.41 &       4.93 &  24.71      &    36.08\\
   3.00  &      1.91 &       5.35 &       1.49 &       1.65 &       5.32 &  22.16      &    37.23 \\
   3.35  &      2.24 &       6.20 &       1.47 &       1.89 &       5.72 &  20.40      &    38.48\\
   3.70  &      2.58 &       7.05 &       1.46 &       2.12 &       6.12 &  19.13      &    39.67 \\
   4.05  &      2.91 &       7.91 &       1.45 &       2.36 &       6.53 &  18.16      &    40.72 \\
   4.40  &      3.25 &       8.77 &       1.44 &       2.60 &       6.94 &   17.40     &   41.62
  \\
\end{tabular}
\end{table*}
\begin{table}[h]
\centerline{y=-600 MeV/c}
\begin{tabular}{|cccccccc|}
$|{\bf q}|$ & $\nu$ & $Q^2 $ & $x_{Bj}$  & $p_{lab} $ & $ s$ & $\sigma_{el}$ & $\sigma_{tot}$\\
$ GeV/c$ & $ GeV$ & $GeV^2/c^2$ &    & $ GeV/c$ & $  GeV^2$ & $ mb$ & $mb$\\
 \hline\hline
 1.30 &        .41 &       1.52 &       1.98 &        .08 &       3.53 &2572.76      &  2572.76 \\
 1.65 &        .65 &       2.30 &       1.90 &        .35 &       3.64 & 129.73      &   129.73  \\
 2.00 &        .92 &       3.14 &       1.81 &        .58 &       3.83 &  59.87      &    59.87  \\
 2.35 &       1.23 &       4.02 &       1.75 &        .81 &       4.08 &  40.76      &    38.37  \\
 2.70 &       1.54 &       4.92 &       1.70 &       1.02 &       4.36 &  32.19      &    35.75  \\
 3.05 &       1.86 &       5.83 &       1.67 &       1.23 &       4.66 &  27.39      &    35.54  \\
 3.40 &       2.19 &       6.75 &       1.64 &       1.44 &       4.98 &  24.35      &    36.20   \\
 3.75 &       2.53 &       7.68 &       1.62 &       1.64 &       5.30 &  22.25      &    37.17  \\
 4.10 &       2.86 &       8.61 &       1.60 &       1.84 &       5.63 &  20.72      &    38.22   \\
 4.45 &       3.20 &       9.55 &       1.59 &       2.04 &       5.97 &  19.56      &    39.24   \\
 4.80 &       3.54 &      10.49 &       1.58 &       2.23 &       6.31 &   18.65     &   40.17
\end{tabular}
\vspace*{1cm}
\caption{Kinematical variables for the inclusive
$D(e,e' )X$ process corresponding to the results
shown in Figs \ref{xshr}-\ref{fyconst}. The various quantities are as
follows:
$|{\bf q}|$,  $\nu$,  and  $Q^2 $, are
 the energy, three-momentum  and four-momentum transfers, respectively;
 $x_{Bj}$ is the Bjorken scaling variable;   $p_{lab}$
is the momentum of the struck nucleon in the final state, defined by
the equation
 $s =
2M^2+2M\sqrt{p_{lab}^2+M^2}$,
where  $ s$ is the Mandelstam variable (cf Eq.(\ref{sman})); finally,
   $\sigma_{el}$ and  $\sigma_{tot}$  are the elastic and total cross sections used in
   the Glauber calculation}
\label{tablitza}
\end{table}

Table.~\ref{tablitza}. C. Ciofi degli Atti....FSI
effects..

\end{document}